\documentclass[10pt]{article}

\usepackage[margin=1in]{geometry}
\usepackage{fancyhdr}
\fancypagestyle{plain}{%
\fancyhf{} 
\fancyfoot[R]{-\,\fontsize{8pt}{8pt}\selectfont\thepage\,-} 

}
\usepackage{amsmath}
\usepackage{amsfonts}
\usepackage{amssymb}
\usepackage{graphicx}
\usepackage{sidecap}
\usepackage{xspace} 
\usepackage{ntheorem}
\usepackage{mathtools}
\usepackage{diagbox}
\usepackage{ntheorem}
\usepackage{multirow}
\usepackage{hyperref}

\usepackage{tensor}

\usepackage{letltxmacro}
\LetLtxMacro{\originaleqref}{\eqref}
\renewcommand{\eqref}{Eq.~\originaleqref}

\newcommand{\dd}{\text d}
\newcommand{\ga}{\alpha}

\newcommand{\equ}[1]{\begin{gather} #1 \end{gather}}
\newcommand{\sfrac}[2]{\mbox{$\frac{#1}{#2}$}}
\newcommand{\order}[1]{\mathcal O\left(#1\right)} 
\newcommand{\abs}[1]{\left\vert#1\right\vert} 
\newcommand{\quads}[1]{\quad #1 \quad}
\newcommand{\qand}{\quad \text{and} \quad}
\newcommand{\maxent}{\textsc{m}ax\textsc{e}nt\xspace}
\newcommand{\ipf}{\textsc{ipf}\xspace}
\newcommand{\mle}{\textsc{mle}\xspace}

\newcommand{\config}{\ga}

\newcommand{\prob}[1]{\mathfrak{#1}}
\DeclarePairedDelimiterX\infdivx[2]{(}{)}{
  #1\;\delimsize\|\;#2
}
\newcommand{\infdiv}{D_\textsc{kl}\infdivx}
\DeclarePairedDelimiterX{\condpX}[2]{(}{)}{%
  #1\;\delimsize\vert\;#2%
}
\newcommand{\condp}{\,\condpX}
\newcommand{\rank}{\text{rank}\xspace}
\newcommand{\Id}{\text{\small 1}\hspace{-3.5pt}\text{1}}
\theoremstyle{break}
\newtheorem{theorem}{Theorem} 
\newtheorem{corollary}{Corollary}[theorem]
\newtheorem{lemma}{Lemma} 

\newcommand{\multidistro}{\text{mult}}
\newcommand{\dimstatespace}{{\abs{\mathcal{A}}}}
\newcommand{\maxentP}{\hat{\prob p}}
\newcommand{\coefM}{\mathbf R}
\newcommand{\coef}{R}

\usepackage[colorinlistoftodos,prependcaption,textsize=tiny]{todonotes}

\begin{document}

\thispagestyle{empty}

\begin{flushright}
\phantom{Version: \today}
\\
\end{flushright}
\vskip .2 cm
\subsection*{}
\begin{center}
{\Large {\bf Entropy-based Characterization\\[1ex]of Modeling Constraints
} }
\\[0pt]

\bigskip
\bigskip {\large
{\bf Orestis Loukas}\footnote{E-mail: orestis.loukas@uni-marburg.de},\,
{\bf Ho Ryun Chung}\footnote{E-mail: ho.chung@uni-marburg.de}\bigskip}\\[0pt]
\vspace{0.23cm}
{\it Institute for Medical Bioinformatics and Biostatistics\\
Philipps-Universität Marburg\\
Hans-Meerwein-Straße 6, 35032 Germany}

\bigskip
\end{center}

\vspace{5cm}
\subsection*{\centering Abstract}
In most data-scientific approaches, the principle of Maximum Entropy (\maxent) is used to a posteriori justify some parametric model which has been already chosen based on experience, prior knowledge or computational simplicity. 
In a perpendicular formulation to conventional model building, we start from the linear system of phenomenological constraints and asymptotically  derive 
the distribution over all viable distributions that satisfy the provided set of constraints. The \maxent distribution plays a special role, as it is the most typical among all phenomenologically viable distributions representing a good expansion point for large-$N$ techniques. 
This enables us to consistently formulate hypothesis testing in a fully-data driven manner. 
The appropriate parametric model that is supported by the data can be always deduced at the end of model selection. 
In the \maxent framework, we recover major scores and selection procedures used in multiple applications and 
assess their ability to capture associations in the data-generating process and identify the most generalizable model. 
This data-driven counterpart of standard model selection demonstrates the unifying prospective of the deductive logic advocated by \maxent principle, while potentially shedding new insights to the inverse problem.

\newpage 
\setcounter{page}{1}
\setcounter{footnote}{0}
%
%

\section{Introduction}

Given phenomenological information in form of marginals, averages or higher moments of features modeling seeks to determine some class of statistical models (i.e.\ probability distributions) that best captures trends or laws in the underlying system that generated the observed data.
Since phenomenological data is by definition a physical invariant of the problem at hand, it is most naturally expressed by the invariant form of the defining set of equations which constrain the space of model distributions to reproduce the observed moments. 
As long as constraints remain of linear form, at least, this poses a well-defined problem eliminating  ambiguities related to reparametrization symmetries in phenomenological information. 

For a long time, it has been known that the distribution which satisfies the problem-defining constraints while maximizing the entropy functional is in a sense the least-biased distribution~\cite{6773024,jaynes2003probability} that incorporates the provided information and nothing more. 
Aside from this conceptually appealing interpretation, the entropy-based approach to modeling offers many benefits of practical significance. 
Given certain information from phenomenology  entropy maximization  uniquely selects \textit{one} model distribution as the most representative among all phenomenologically viable distributions offering a clear deductive reasoning to modeling. 
At the same time, the \maxent logic  combinatorially assigns probabilities to any other distribution in the most intuitive way, namely according to its ``distance'' from the provided set of observations.
As this point appears to be of fundamental importance in the \maxent logic, we start our more formal parameter-agnostic exploration by investigating the distribution over phenomenologically viable distributions.

Since entropy measures  information flowing into our model, the \maxent principle applied on the linear system of phenomenological constraints turns also classification tasks into clear-cut deductive procedures.
In this work, we explain how to unambiguously define model architectures and directly count the physical degrees of freedom defining model dimensionality in order to ultimately perform selection of significant constraints that can be distinguished from sampling noise. 
Furthermore, we examine how large-order expansions of 
information-theoretic measures and benchmark scores 
can be meaningfully evaluated, when performed around the distribution of maximum entropy.  

More specifically,  we rigorously show that expansions in the fluctuations of probabilities around their \maxent estimate  at large sample size $N$ result in leading Gaussian kernels with terms that are quadratic in the fluctuating degrees of freedom. 
Such formal observation justifies  writing discrete distributions over distributions into multivariate normal distributions in the fluctuations around the \maxent saddle point.
Asymptotically, this suggests the integral representation of expectation sums over viable distributions which can be in turn evaluated using standard large-$N$ localization techniques. Eventually, we arrive at  hypothesis testing in the \maxent world and suggest significance level(s) compatible with the large-$N$ expansion. In addition, we review standard modeling scores outlining their meaning in the present framework.

Under the unifying umbrella of \maxent reasoning, we test accuracy of (hyper)graph reconstruction and generalizability of models (i.e.\ \maxent distributions satisfying sets of constraints) trained on synthetic data from an effective Ising magnet. 
Taking $N$ large verifies the theoretically anticipated estimates of error bounds. Most crucially, it reveals the universality of many measures which is expected to manifest at large order: Information-theoretic ``distances'' between distributions asymptotically depend only on the  number of physical degrees of freedom and not on further details of the phenomenological problem at hand. In particular, most model selection procedures seem to merge into a universal rule that accurately detects the asymptotic structure.

All in all, the constraint optimization of entropy gives a robust definition of the notion of a \textit{model} assigning to any distinct set of consistent constraints its \maxent distribution. 
At an operational level, the simplicity and flexibility of directly working with physical degrees of freedom, i.e.\ probabilities,  could help efficiently tackle inverse problems in data science for which parameters (lacking microscopic justification) often lead to distractingly cumbersome inference. 
In addition, it enables the large-$N$ estimation of important information-theoretic metrics, such as the expected entropy, training and test errors. 
Under a unifying framework, the \maxent principle seems to naturally provide the tools to consistently perform application-adaptive selection of the most suitable model reflecting underlying mechanisms in a fully data-driven  fashion.

\section{Phenomenological constraints}
\label{sc:PhenoProblem}

Any phenomenological problem is related to some probability distribution over physical realizations.  
In this work, we concretely investigate statistical systems of categorical features each assuming a distinct number of states. All admissible realizations of these features comprise the space of microstates $\mathcal A$.
A distribution $\prob p$ over $\mathcal A$ can be thought of as an $\dimstatespace$-dimensional vector living on the probability simplex $\mathcal P$ which assigns a probability to each microstate.
Modeling seeks to determine some optimal distribution(s) in $\mathcal P$ that satisfies $D\geq1$ phenomenological constraints (one is the normalization condition in $\mathcal P$) which we denote by a vector $\widehat{\prob m}$.

The most obvious and easy to handle type of constraints is of linear form,
\equ{
\label{eq:LinearSystem}
\sum_{\ga=1}^\dimstatespace \coef_{a\ga}\, \prob p^\ga \overset{!}{=} \widehat m_a\quads{\text{for}}a=1,\ldots D
~,
}
based on an architecture matrix $\coefM$ with each row representing a constraint and columns corresponding to the microstates in $\mathcal A$. 
In the following, we take the architecture matrix to be of full row rank $D=\rank\mathbf \coefM$ directly specifying the number of independent constraints.
In realistic applications however, a redundant set of linear constraints summarized by a coefficient matrix $\mathbf C$ acting on $\prob p$  is often provided, instead. 
Whenever the phenomenological input concretely consists of a set of marginal relative frequencies, $\mathbf C$ becomes a binary matrix dictating the relative frequencies of microstates to be summed over to obtain marginal relative frequencies in the empirical set (i.e.\  each row of the coefficient matrix describes a marginal constraint sum).
To obtain a robust model definition it is instructive to transform 
the coefficient matrix $\mathbf C$ to its reduced row-echelon form.  Dropping any all-zero row corresponding to redundant constraints, one then obtains the architecture matrix $\coefM$ which uniquely characterizes the linear system at hand.

Noting in particular  that there are by construction $D$ columns in $\coefM$ with all-zeros but one (the leading) 1,
it can be easily seen  that normalization of $\prob p\in\mathcal P$ implies
\equ{
\label{eq:Normalization}
\sum_{a=1}^{D} \coef_{a\ga} = 1 \quad\forall\, \ga=1,\ldots,\dimstatespace
\qand
\sum_{a=1}^{D} \widehat m_a = 1
~.
} 
The solution set of linear system \eqref{eq:LinearSystem}, which is unambiguously defined by architecture matrix $\coefM$ and inhomogeneous vector $\widehat{\prob m}$,  induces an equivalence class in $\mathbb R^\dimstatespace$.
To see this first note that the empirical distribution $\prob f\in\mathcal P$ which summarizes the observed relative frequencies of all microstates in $\mathcal A$ always satisfies the coupled equations, thus being member of the equivalence class which we denote by $[\prob f]_{\coefM}$ in the following. By construction, any distribution $\prob p\in[\prob f]_{\coefM}$ shares the same statistics $\widehat{\prob m}$ as the empirical distribution.
Whenever only  normalization constraint is specified as an all-one row in $\coefM$, the equivalence class covers the full simplex, $[\prob f]_{(1,\dots,1)}=\mathcal P$, whereas in the other extreme case where the architecture matrix becomes the identity, all frequencies are learned by heart,  so that $[\prob f]_{\Id} = \lbrace \prob f\rbrace$.

In a general setup, there might exist microstates whose probability is fixed in the problem statement to some (non)-finite empirical value. 
If some empirical marginal is known to be zero by logical deduction, for example, all the associated probabilities  appearing in this marginal sum are necessarily set to zero due to the non-negativity on the simplex $\mathcal P$.
In the following, we concentrate on all constraints and associated  microstates   that are not a priori determined by some empirical or logical argument removing any 
probabilities that are learned by heart from empirical data. The latter can be trivially reinstated at the end of model selection without modifying conclusions. In other words, we understand $\mathcal A$ to only include those microstates with $\prob p_\ga>0$ that retain some probabilistic  interest for model building.

\section{Entropy concentration}
\label{sc:ConcentrationTH}
Mathematically, the notion of an optimal solution amounts to selecting a target functional to be extremized over the equivalence class associated to the system of phenomenological constraints. 
Choosing~\cite{shore1980axiomatic,paris1990note,csiszar1996maxent} as a  measure of uncertainty Shannon's differential entropy,
\equ{
H[\prob p] = - \sum_{\ga=1}^\dimstatespace \prob p_\ga \log \prob p_\ga\quads{\text{with}} \prob p\in\mathcal P~,
}
we are naturally led to the principle of maximum entropy (\maxent) under linear constraints \eqref{eq:LinearSystem}. In this way, we ensure that only the information specified by the linear system in \eqref{eq:LinearSystem} enters into the model distribution and nothing more.
If no constraints were specified besides the dimensionality of $\mathcal A$, then the distribution of maximum entropy would equal the uniform distribution $\prob u_\ga=\dimstatespace^{-1}$ as a manifestation~\cite{keynes1921chapter} of having the same degree of ignorance about any two microstates in $\mathcal A$. 
In the other extreme scenario that all $\dimstatespace$ empirical frequencies were constrained, \eqref{eq:LinearSystem} would unambiguously fix the \maxent distribution to the empirical distribution $\prob f$, as previously discussed. 
Most interestingly, we are concerned with all those cases were $\coefM$ is column-rank deficit thus leading to a non-trivial equivalence class $[\prob f]_{\coefM}$ from which the \maxent principle selects a representative distribution that we henceforth denote by $\maxentP$. As established in~\cite{loukas2022categorical} the \maxent distribution always exists and is unique for any consistent set of linear constraints.

Suppose we are now given a dataset with $N$ records. From elementary combinatorics, we know that the multinomial coefficient (r.h.s.\ gives the large-$N$ expansion)
\equ{
\label{eq:MultinomialCoefficient}
W[N\prob p] = \frac{N!}{\prod_{\ga=1}^\dimstatespace (N\prob p^\ga)!}
=
\exp\left\lbrace N H[\prob p]
-\frac{\abs{\mathcal A}-1}{2} \log N
+ \order{1}
\right\rbrace
}
counts the possible ways to separate those $N$ individuals into $\dimstatespace$ teams such that the counts $N \prob p$ are reproduced. 
As we have no way to distinguish among realizations of the $N \prob p$ counts, we should assign equal probabilities to them. 
Under this thermodynamic assumption, member distributions in the equivalence class formally distribute according to
\equ{
\label{eq:DistributionOfDistributions}
p\condp{\prob p}{N\prob f,\coefM} = \left(\sum_{\prob p'\in[\prob f]_\coefM} \exp\left\{ NH[\prob p']\right\}\right)^{-1}
 \exp\left\{ NH[\prob p]\right\}
~,
}
at sample sizes $N\gg1$. Incorporating model-dependent effects of $\order{1}$ in the large-$N$ expansion of \eqref{eq:MultinomialCoefficient} would result in $\order{1/N}$ effects in any relevant expectation over the distribution of distributions in the equivalence class.
At sufficiently large $N$ we may thus neglect subleading corrections in Stirling's formula used to expand factorials and directly start from \eqref{eq:DistributionOfDistributions}.

In particular, the ratio $W[N\maxentP]/W[N\prob p]$ for any distribution $\prob p\in[\prob f]_{\coefM}$ measures the relative abundance of the \maxent distribution over another member in the equivalence class.
The large-$N$ expansion of \eqref{eq:MultinomialCoefficient} leads to 
\equ{
\label{eq:Motivation:ExponentialDominanance}
\log\frac{W[N\maxentP]}{W[N\prob p]} = N\left(H[\maxentP] - H[\prob p]\right) + \order{1}
~.
}
Taking the $N\rightarrow\infty$ limit, we recognize~\cite{jaynes1968prior} that the \maxent distribution can be realized in more ways compared to any other distribution in $[\prob f]_{\coefM}$, as the entropy difference will be by definition  positive.
Consequently, $\maxentP$ is the distribution we should overwhelmingly  anticipate given the empirical constraints, the latest asymptotically, since any other distribution would be exponentially suppressed, justifying our choice of target functional.
Intuitively, the dominating role of $\maxentP$ over any other distribution in the equivalence class shows that the \maxent distribution is in a sense \textit{typical} of this set. 

In order to prove the main and subsequent theorems in this paper we shall need a fairly general  
\begin{lemma}\label{lm:Fluctuations}
Let $\maxentP$ be the \maxent distribution in equivalence class $[\prob f]_{\coefM}$ induced by architecture $\coefM$ on data described by empirical distribution $\prob f\in\mathcal P$. For any distribution $\prob p\in[\prob f]_{\coefM}$ with linearized fluctuations 
\equ{
\label{eq:LinearizedFluctuations}
\prob p_\ga = \maxentP_\ga  
+
\sqrt{\frac{\maxentP_\ga}{N}}\, \boldsymbol\pi_\ga
}
around the \maxent distribution in the equivalence class, the large-$N$ expansion of the entropy difference 
\equ{
\label{eq:EntropyDifference_LargeN}
2N\left(H[\maxentP] - H[\prob p]  \right) = 
\sum_{\ga=1}^{\vert\mathcal A\vert}  \boldsymbol\pi_\config^2
 + \order{N^{-1/2}}
}
results into a leading  term which is quadratic in the fluctuations $\boldsymbol\pi\in\mathbb R^\dimstatespace$.
\end{lemma}

\paragraph{Proof}
As long as\footnote{The case of $\maxentP_\ga=0$ exhibits no probabilistic fluctuations and does not contribute to our formulas. Formally, this is manifested by taking $0\log0=0$ in any information-theoretic measure.} $\prob p_\ga\neq0$, parametrizing fluctuations according to \eqref{eq:LinearizedFluctuations} does not degenerate. 
Generically, normalization on the simplex $\mathcal P$ means 
\equ{
\label{eq:Normalization_Fluctuations}
\sum_{\ga=1}^\dimstatespace \boldsymbol\pi_\ga\sqrt{\maxentP_\ga}=0
~.
}
Furthermore,  invoking a well-known fact~\cite{csiszar1975divergence,amari2000methods} from information theory about \maxent distributions $\maxentP$,
\equ{
\label{eq:BaseLemma}
\sum_{\ga=1}^\dimstatespace \left(\prob p_\ga - \maxentP_\ga \right) \log\maxentP_\ga = 0 \quad \forall\,\prob p\in[\prob f]_{\coefM} 
~,
}
we recognize that
\equ{
\label{eq:BaseLemma_Fluctuations}
\sum_{\ga=1}^\dimstatespace\boldsymbol\pi_\ga\sqrt{\maxentP_\ga}\log\maxentP_\ga=0
~,
}
whenever the \maxent distribution in $[\prob f]_{\coefM}$ is chosen as  expansion point.
Combining \eqref{eq:Normalization_Fluctuations} with \eqref{eq:BaseLemma_Fluctuations} in the Taylor expansion of the entropy difference results into \eqref{eq:EntropyDifference_LargeN}.
The factor of $\sqrt{N}$ in parametrization of \eqref{eq:LinearizedFluctuations} is chosen~\cite{geyer2013asymptotics} so that the quadratic term in \eqref{eq:EntropyDifference_LargeN} scales as $\order{1}$ in the large-$N$ expansion ensuring the proper asymptotic limit by suppressing higher-order corrections.

We are now in a position to state 
\begin{theorem}
\label{th:Concentration}
Asymptotically, the  scaled difference $2N(H[\maxentP] - H[\prob p])$ of  the entropy of any distribution $\prob p\in[\prob f]_{\coefM}$ from the maximum entropy in the equivalence class induced by $\coefM$ on data described by $\prob f$ is  distributed as chi-squared with 
$\abs{\ker \coefM}=\dimstatespace-D$ degrees of freedom, in short as $\chi_{\dimstatespace-D}^2$.
For any positive $\delta$ the likelihood to find a fraction of distributions in $[\prob f]_\coefM$ having entropies in the range  
\equ{
H[\maxentP] - \delta \leq H[\prob p] \leq H[\maxentP]
\nonumber
}
is accordingly given by the cumulative distribution $F_{\dimstatespace-D}(2N\delta)$ of $\chi_{\dimstatespace-D}^2$ distribution. 
\end{theorem}

\paragraph{Proof}
At sufficiently large $N$, 
we are interested to evaluate the fraction of distributions in the equivalence class having entropies in a range $\delta>0$ below \maxent $H[\maxentP]$, i.e.\ in the region
\equ{
\mathcal R = \left\lbrace \prob p\in[\prob f]_\coefM \,\big\vert\, H[\prob p]\geq H[\maxentP] - \delta\right\rbrace
~.
}
Formally using the distribution $p$ over distributions introduced in \eqref{eq:DistributionOfDistributions} this fraction is written as
\equ{
\label{eq:FractionDEF}
F_k(2N\delta) = \sum_{\prob p\in\mathcal P} \Id_\mathcal{R}\,p\condp{\prob p}{N\prob f,\coefM}
=
\left(\sum_{\prob p'\in[\prob f]_\coefM} \exp\left\{ -N(H[\maxentP] - H[\prob p'])\right\}\right)^{-1}
\sum_{\prob p\in\mathcal R} \exp\left\{ -N(H[\maxentP] - H[\prob p])\right\}
~,
}
($\Id_\mathcal{R}$ denotes the indicator function), which is clearly governed by the entropy difference $H[\maxentP] - H[\prob p]$ of each member distribution from \maxent.
For the purpose of evaluating $p$ as an expansion in powers of $1/N$, we set
\equ{
\label{eq:FluctuationsDEF}
\boldsymbol\pi_\config = \sqrt{\frac{N}{\maxentP_\ga}} \left[\prob p_\config - \maxentP_\config\right] 
\quad\forall\,\ga=1,\ldots,\dimstatespace
}
such that $\boldsymbol\pi_\config=\order{\sqrt N}$. 
Clearly,  Lemma~\ref{lm:Fluctuations} applies for $\boldsymbol\pi$, so that the governing entropy difference results in a quadratic term as in \eqref{eq:EntropyDifference_LargeN}.
On account of linear system \eqref{eq:LinearSystem},
fluctuations around any reference distribution in the equivalence class satisfy homogeneous system
\equ{
\label{eq:HomogeneousConstraints}
\sum_{\config=1}^\dimstatespace \left(\coef_{a \config}\sqrt{\maxentP_\ga}\right) \boldsymbol\pi_\config = 0 ~~\forall\,  a=1,\ldots,D 
~.
}
Due to these homogeneous constraints any $\boldsymbol\pi\in\mathbb R^{\vert\mathcal A\vert}$ lives 
in the $(\dimstatespace-D)$-dimensional cone spanned by the kernel of architecture matrix $\coefM$ after column-wise  rescaling by $\sqrt{\maxentP_\ga}>0$.
At an operational level, the quadratic sum in \eqref{eq:EntropyDifference_LargeN} expressed in any valid coordinate system parameterizing the kernel of the rescaled architecture matrix can be  diagonalized with unity-det Jacobian. 
To this end, we can write the fluctuations in \eqref{eq:LinearizedFluctuations} around \maxent solution as  
\equ{
\boldsymbol\pi_\ga =  \sum_{i=1}^{\dimstatespace-D} x_i X_{i\ga} 
~,
}
where $\mathbf X$ orthonormally encodes the kernel space, i.e.\ for $a=1,\dots,D$ and $i,j=1,\dots,\dimstatespace-D$ it is 
\equ{
\label{eq:OrthonormalKernel}
\sum_{\ga=1}^\dimstatespace \left(\coef_{a \config}\sqrt{\maxentP_\ga}\right) X_{i\ga}=0 
\qand
\sum_{\ga=1}^\dimstatespace X_{i\ga} X_{j\ga}=\delta_{ij}
~.
} 
Hence, the leading quadratic term in  
the entropy difference becomes
\equ{
\sum_{\ga=1}^{\vert\mathcal A\vert}  \boldsymbol\pi_\config^2
= \sum_{i=1}^{\dimstatespace-D} x_i^2 \,\sim\, \chi^2_{\dimstatespace-D}
~,
}
readily 
following  a chi-square distribution with $\dimstatespace-D$ degrees of freedom.

Exploiting the radial symmetry  of the large-$N$ expansion we subsequently introduce spherical coordinates such that any $\prob p\in[\prob f]_\coefM$ living on the $(\dimstatespace-D - 1)$\,--\,sphere with squared radius $\chi^2=\sum_{i=1}^{\dimstatespace-D}x_i^2$ has leading entropy 
\equ{
\label{eq:EntropicRadius}
H[\prob p] = H[\maxentP] - (2N)^{-1} \chi^2
~.
}
Notice that the uniqueness of the optimal solution unambiguously fixes the origin of the chosen spherical coordinate system 
to correspond to the \maxent distribution in the given equivalence class. 
Conveniently, the spherical parametrization allows us to write the integral representation of \eqref{eq:FractionDEF} after performing the angular integration,
\equ{
\label{eq:radialIntegral}
F_{\dimstatespace-D}(2N\delta) = 
\frac{1}{2^{\frac{\dimstatespace-D}{2}-1}\, \Gamma\left(\frac{\dimstatespace-D}{2}\right)}
\int_0^{\sqrt{2N\delta}} \dd \chi\, \chi^{\dimstatespace-D-1} \, e^{-\chi^2/2}
~,
}
as the cumulative distribution of chi squared.
This concludes the proof.
\\

Based on radial symmetry, Theorem~\ref{th:Concentration} provides~\cite{Rosenkrantz1989} a neat geometrization of the distribution over distributions \eqref{eq:DistributionOfDistributions} in the equivalence class. 
Around the \maxent distribution $\maxentP$ in a $\abs{\ker \coefM}$-dimensional space, any spherical shell of radius $\sqrt{2N \Delta H}$ traces over all those distributions $\prob p\in[\prob f]_{\coefM}$ whose entropy differs from $H[\maxentP]$ by the same amount $\Delta H$. 
Clearly, the interval within which the entropy of a specified portion of distributions is to be found becomes narrower with larger $N$, showing that the \maxent distribution has an entropy more and more representative of the whole equivalence class. Hence, the allowed by $\coefM$ entropies \textit{concentrate} around the \maxent in a geometric sense. A posteriori, this formally justifies our choice of entropy (which in the present setting rather emerges at large-$N$ than being postulated) as a target functional for selecting a model distribution.

\subsection{Calculating the optimal solution}
\label{sc:IterativeMethods}
The preceding discussion about the concentration of entropies around the \maxent and the application of chi-squared statistics presupposes that the distribution of maximum entropy has already been determined. Of course, one can parametrize the solution space of linear system \eqref{eq:LinearSystem} and try to maximize analytically the entropy. For small dimensionality of $\ker\mathbf R$, the solution could have a neat closed form. In other setups, there might exist some perturbative method to proceed.

In a reverse approach, we directly incorporate the optimization objective into the phenomenological system to obtain a non-linear system of coupled equations, which 
can be iteratively solved via Newton-Raphson method.
Given architecture matrix $\coefM$ and empirical inhomogeneous vector $\widehat{\prob m}$, we start from the uniform distribution $\prob p^{(0)}=\prob u$. 
Upon using the exponential form~\cite{csiszar1975divergence} of \maxent distribution in absence of zero-probability microstates,
\equ{
\label{eq:ExponentialForm}
\maxentP_\ga = \exp \left(\sum_{a=1}^D \hat\theta_a\coef_{a\ga}\right)
~,
}
for some vector $\hat{\boldsymbol\theta}\in\mathbb R^D$ of Lagrange multipliers,
the update rule for Newton estimate of  the \maxent solution to system \eqref{eq:LinearSystem} at the $n$-th step reads
\equ{
\prob p_\ga^{(n+1)} = \prob{p}_\ga^{(n)} \exp\left\lbrace - \sum_{a,b=1}^D \coef_{a\ga} ((\mathbf J^{(n)})^{-1})_{ab} \left(m_b^{(n)}-\widehat{m}_b\right) \right\rbrace 
\quad\text{for }\ga=1,\ldots,\dimstatespace
}
in terms of  the previously estimated empirical moment
\equ{
m_a^{(n)} = \sum_{\ga=1}^\dimstatespace \coef_{a\ga} \,\prob p^{(n)}_{\ga}
\quads{,} a=1,\ldots,D
}
and Jacobian matrix 
\equ{
J^{(n)}_{ab} = \sum_{\ga=1}^\dimstatespace \coef_{a\ga}\, \prob{p}^{(n)}_\ga\, \coef_{b\ga}
\quads{,} a,b=1,\ldots,D
~.
}
The latter matrix is invertible, as it can be easily seen by noting that vanishing of its quadratic form for any vector $\prob u\in\mathbb R^D$ implies 
\equ{
\label{eq:InvertibleJacobian}
0\overset{!}{=}\sum_{a,b=1}^D u_a J^{(n)}_{ab} u_b = \sum_{\ga=1}^\dimstatespace  \prob p^{(n)}_\ga  \left(\sum_{a=1}^D u_a \coef_{a\ga}\right)^2 
\quads{\Rightarrow}
\sum_{a=1}^D u_a \coef_{a\ga} = 0 
\quads{\Rightarrow}
u^a = 0 \quad\forall a=1,\ldots D~,
}
since all $\prob p^{(n)}_\ga>0$ (having excluded structural zeros whose Newton estimate becomes trivially zero) while the rows of $\coefM$ are per assumption linearly independent.  
As customary, Newton method converges fast to the unique \maxent distribution $\maxentP$, at the cost of a $D\times D$ matrix inversion at each iteration step.

There exist many variations of the main iterative scheme.  Suppose that instead of $\coefM$ a binary matrix $\mathbf C$    which describes a  possibly redundant set of $M$ marginal constraints is used. Cycling through each marginal constraint $a=1,\ldots M$ to iteratively update probabilities via one-dimensional Newton method (as the multidimensional Newton method would be plagued by the singularity of redundancies) gives the rule
\equ{
\label{eq:IPF}
\prob p^{(nM+a)}_\ga = \prob p^{(nM+a-1)}_\ga \exp\left\lbrace C_{a\ga} \left(\frac{\widehat{m}_a}{m^{(nD+a-1)}_a} -1\right)\right\rbrace 
\approx 
\prob p^{(nM+a)}_\ga \left(\frac{\widehat{m}_a}{m^{(nM+a-1)}_a}\right)^{C_{a\ga}}
~,
}
which is determined by the ratio of empirical marginals to the running estimate of model marginals
\equ{
m^{(nM+a-1)}_a = \sum_{\ga=1}^\dimstatespace  C_{a\ga} \prob p^{(nM+a-1)}_\ga
~.
}
One can show~\cite{csiszar1975divergence,ireland1968contingency} that the Taylor-expanded r.h.s.\  of \eqref{eq:IPF} always converges to the \maxent solution. This approach leads to the algorithm of iterative proportional fitting~\cite{bishop2007discrete}.

\section{Constraint selection}
\label{sc:ConstraintSelection}

In this section, we exploit the logic advocated in Section~\ref{sc:ConcentrationTH} as our starting point
in order to motivate a fully data-driven selection of phenomenological constraints.
In most cases, we are given summary statistics of an empirical distribution, i.e.\ a set of phenomenological constraints summarized by $\widehat{\mathbf C}$  whose invariant form $\widehat{\coefM}$ induces an equivalence class $[\prob f]_{\widehat\coefM}\subseteq\mathcal P$, as outlined in Section~\ref{sc:PhenoProblem}. This robustly defines the notion of a \textit{model} in a purely data-driven  
fashion. 
In addition, a  chi-square distribution  asymptotically assigns probabilities to all member distributions in the equivalence class based on their entropy relative to \maxent. With increasing sample size $N$, an increasing number of distributions has entropies concentrating around \maxent, making $\maxentP$ typical of $[\prob f]_{\widehat\coefM}$ and hence a good candidate of a model distribution. 

\subsection{Hypothesis testing}
\label{sc:HypothesisTesting}

\paragraph{The hyper MaxEnt.}

If the given architecture matrix $\widehat{\coefM}$ defined all our knowledge about the system under investigation,
we first ask how likely it would be to observe the empirical distribution $\prob f$
whose entropy differs from the \maxent by  
\equ{
\label{eq:empirical_p_value}
\hat\delta=H[\maxentP]-H[\prob f]
~.
}
On account of Theorem~\ref{th:Concentration},
the asymptotic likelihood to observe the empirical distribution --\,if the $D$ constraints in $\widehat{\coefM}$ are all we know/trust from the data\,-- is naturally    
given as the $p$-value under cumulative $\chi^2_{\dimstatespace-D}$ distribution to record a value $2N\hat\delta$ or more extreme. If the empirical $p$-value lies above a pre-specified threshold, we could argue in favor of the \maxent distribution $\maxentP$ as capturing all essential information, any discrepancy to $\prob f$ attributed to  mere statistical noise. In the opposite case of a $p$-value below the threshold, we would have to deduce that the provided constraints were not sufficient to capture mechanisms of the data-generating process, $\maxentP$ systematically failing to adequately characterize $\prob f$.

In principle, full access to  empirical distribution $\prob f$ makes a classification of all possible --\,given the microstate space $\mathcal A$\,-- architectures $\widehat{\coefM}$ and corresponding generalized moments $\widehat{\prob m}$ amenable. Any set of constraints which fails to capture the systematics of the data-generating process, so that the observed data $\prob f$ becomes very atypical in the induced equivalence class,  is immediately vetoed. 
In the spirit of Occam's razor, we select the simplest model with the highest entropy which still makes the empirical dataset likely (empirical $p$-value above the threshold) as our starting point. 
We shall call the model incorporating all significant constraints --\,i.e.\ constraints whose removal leads to significantly low $p$-values\,-- and nothing more, the \textit{hyper}-\maxent.

\paragraph{Likelihood ratio test.}
For those constraints whose removal does not bring any significant result, as far as the empirical entropy difference $H[\maxentP]-H[\prob f]$ is concerned, we could proceed by relative comparison of constraints according to 
\begin{theorem}
\label{th:Concentration2}
Given is a base architecture $\widehat\coefM$ and a more complex $\widehat\coefM'$ which implies the former. 
Asymptotically, the  scaled difference $2N(H[\maxentP] - H[\maxentP'])$ of  \maxent in $[\prob f]_{\widehat{\coefM}'}\subset[\prob f]_{\widehat{\coefM}}$ from  \maxent in $[\prob f]_{\widehat{\coefM}}$ is  distributed as chi-squared with  
$\widehat D' - \widehat D$ degrees of freedom.
\end{theorem}
\noindent
Evidently, Theorem~\ref{th:Concentration2} reduces to Theorem~\ref{th:Concentration} using as most complex model architecture $\widehat{\coefM}^\prime = \Id$ with $\widehat D'=\dimstatespace$ that learns by heart.

\paragraph{Proof} 
This theorem describes a frequently encountered task in model building, to namely quantify the suitability of a simpler set of constraints $\widehat{\coefM}$ over a more stringent  $\widehat{\coefM}'$ which implies the former. 
At the level of architecture matrices, such nested models are related via a linear map
\equ{
\label{eq:LRT:NestedModels}
\widehat{\coef}_{a\ga} = \sum_{a'=1}^{\widehat D'} T_{aa'} \widehat{\coef}^\prime_{a'\ga}
\quads{,} a=1,\dots,\widehat D
}
for some transformation matrix $\mathbf T$ of $\rank\mathbf T=\widehat D$ (otherwise $\widehat\coefM$ would be row-rank deficit).
In our framework, both models are unambiguously characterized by \maxent distributions $\maxentP$ and  $\maxentP'$, respectively. 
Clearly, the form \eqref{eq:ExponentialForm} of \maxent distribution $\maxentP'$ automatically accommodates the reference distribution of the simpler model, so that there always exists vector $\hat{\boldsymbol\theta}$ with $\maxentP'=\maxentP$ given empirical data.

In a designer setup, different datasets that respect the reference constraints in $\widehat\coefM$ are produced. 
By definition, a \maxent{ist} trusting $\widehat{\coefM}$ would always conclude out of all those datasets that $\maxentP$ is the appropriate distribution.
On the other hand,  a modeler trusting $\widehat{\coefM}'$ would compute an a priori different \maxent distribution $\maxentP'$  depending on the received dataset, but respecting the reference constraints from $\widehat{\coefM}$.
Writing $\maxentP'$ as in \eqref{eq:LinearizedFluctuations} this translates into  fluctuations $\boldsymbol\pi$  around the reference \maxent distribution $\maxentP$ satisfying projection conditions 
\equ{
\label{eq:LRT:refProjection}
\sum_{\ga=1}^\dimstatespace\left(\widehat{\coef}_{a\ga}\sqrt{\maxentP_\ga}\right)\boldsymbol\pi_{\ga} = 0 \quad\forall\, a=1,\dots, \widehat D
~.
} 
We are interested in all those cases that the latter modeler would select a different distribution whereas the first would insist on $\maxentP$. To properly quantify this discrepancy we need to filter out redundancies whenever different datasets lead to the same model distributions, as they differ in constraints that none of the two modelers can see so that both insist on their respective \maxent models.  In other words, fluctuations from $\maxentP'$  around $\maxentP$ must be perpendicular to the kernel of $\widehat{\mathbf R}'$. 
Hence, its linearized fluctuations  must look like
\equ{
\boldsymbol\pi_\ga=\sum_{a'=1}^{\widehat D'}x_{a'}\widehat{\coef}'_{a'\ga}
~.
}
Reference  projection \eqref{eq:LRT:refProjection} then constrains the vector $x\in\mathbb R^{\widehat D'}$ via 
\equ{
\sum_{a'=1}^{\widehat D'} T_{aa'} x_{a'} = 0
\quad\forall\,a=1,\dots,\widehat D~,
}
upon using \eqref{eq:LRT:NestedModels} 
Hence, $x$ lies in the kernel of $\mathbf T$ which is of dimension $\widehat D'-\widehat D$, so that in total
\equ{
\boldsymbol\pi_\ga=\sum_{i=1}^{\widehat D'- \widehat D}y_i \sum_{a'=1}^{\widehat D'}Y_{ia'}\widehat{\coef}'_{a'\ga}
~,
}
where $\mathbf Y$ orthonormally spans the kernel of $\mathbf T$.
Analogously to the proof of Theorem~\ref{th:Concentration}, this form of   fluctuations with $\boldsymbol\pi\cdot\boldsymbol\pi=y\cdot y$ directly leads to the claimed chi-squared distribution by applying Lemma~\ref{lm:Fluctuations}.

In the large-$N$ expansion, the multinomial distribution for any two distributions $\prob f,\prob q\in\mathcal P$,
\equ{
\label{eq:multidistro:expansion_LargeN}
\log\multidistro(N\prob f;\prob q) = -N \infdiv{\prob f}{\prob q} -\frac{\abs{\mathcal A}-1}{2} \log N + \order{1}~,
}
is governed\footnote{The same applies for the multivariate hypergeometric distribution, as long as the population size and population counts are much larger than the size and frequencies of samples from this population.} by the \textsc{kl} divergence of distribution $\prob q$ from empirical $\prob f$.  
In particular, the log-likelihood ratio to leading order in $N$ upon using \eqref{eq:BaseLemma} equals 
\equ{
\log\frac{\multidistro(N\prob f;\maxentP')}{\multidistro(N\prob f;\maxentP)}  = N \left(H[\maxentP] - H[\maxentP']\right) + \order{1} ~,
}
i.e.\ the observed entropy difference between \maxent distributions of the more complex to simpler model. Hence, we recover the likelihood ratio test~\cite{silvey1975statistical} (see also Lagrange multiplier test~\cite{doi:10.1080/00031305.1982.10482817}) in the \maxent framework, 
where $\maxentP$ is preferred, whenever it is sufficiently descriptive of all those distributions satisfying the additional constraints in $\widehat{\coefM}'$ and nothing more.  Conversely, we can formulate
\begin{corollary}
For any two sets of constraints described by architectures $\widehat{\coefM}$ and $\widehat{\coefM}'$ where the former is implied by the latter, the likelihood ratio of an observed dataset of sample size $N$
is governed by the entropy difference $\hat\delta=H[\maxentP]-H[\maxentP']$ between the \maxent{}s in the induced equivalence classes. 
The \maxent distribution  $\maxentP\in[\prob f]_{\widehat{\coefM}}$ is rejected as a good model distribution in favor of $\maxentP'\in[\prob f]_{\widehat{\coefM}'}$ at $\alpha$ significance level, whenever $1-F_{\widehat D'-\widehat D}(2N\hat\delta) < \alpha$. 
\end{corollary}

\subsection{Information criteria}

\paragraph{Maximum likelihood and BIC score.}
In the equivalence class induced by architecture matrix $\widehat\coefM$ on counts $N\prob f$, the probability density $\rho$  of observing  entropy difference $H[\maxentP] - H[\prob p]\equiv r^2/2$ is given according to 
\equ{
\label{eq:ChiDensity}
\rho\condp{\prob p}{N\prob f,\widehat{\coefM}}\,r^{\dimstatespace-\widehat D-1}\dd r \,\dd\Omega = N^{\frac{\dimstatespace-\widehat D}{2}} e^{-N r^2/2}\,r^{\dimstatespace-\widehat D-1}\dd r \,\dd\Omega
~,
}
after redefining the entropic radius $\chi = \sqrt{N} r$ in \eqref{eq:EntropicRadius} to extract the large-order scaling which remains constant throughout the equivalence class.
Taking the logarithm of \eqref{eq:ChiDensity} evaluated at the empirical entropy difference, 
\equ{
\log  \rho\condp{\prob f}{N\prob f,\widehat{\coefM}} = - N \left(H[\maxentP]-H[\prob f]\right) + \frac{\dimstatespace-\widehat D}{2} \log N =  \sfrac12\dimstatespace \log N + NH[\prob f] - \sfrac12\widehat{\textsc{bic}} ~,
}
we recover from the model-dependent part the Bayesian Information Criterion (in short \textsc{bic} score), first introduced in~\cite{10.1214/aos/1176344136}.
In the spirit of \mle, the set of constraints must be selected which make the data most likely to be observed.
One thus needs to choose the architecture $\widehat \coefM$ which assigns highest weight to the observed dataset $\prob f$ or in other words whose penalized empirical entropy difference maximizes the density $\rho$ in the induced equivalence class $[\prob f]_{\widehat \coefM}$, conversely minimizing the \textsc{bic} score.

\paragraph{The expectation of entropy and AIC score.}
Another way to characterize an equivalence class and assess the quality of modeling is via expectations over member distributions $\prob p\in[\prob f]_{\widehat\coefM}$. For instance, the expected entropy in $[\prob f]_{\widehat{\coefM}}$ equals 
\equ{
\label{eq:ExpectedEntropy}
N \langle H[\prob p]\rangle
 = 
N H[\maxentP] -   \langle N(H[\maxentP] - H[\prob p])\rangle 
=
NH[\maxentP] - \frac{\dimstatespace-\widehat D}{2}
~,
}
using the entropic radius from \eqref{eq:EntropicRadius} and taking the expectation over chi-squared density. Simplicity (lower dimension $\widehat D$) seems to dictate the expected entropy to lie further away from \maxent, cf.\ chi-squared probability density spreading out with increasing degrees of freedom. On the other hand, goodness of fit translates into the expected entropy approaching the empirical one. Combining both heuristic objectives we require the expected entropy to be closer to empirical $H[\prob f]$ than to \maxent.
In other words, we could look for the minimum of the difference of entropy differences 
\equ{
N\left(\langle H[\prob p]\rangle - H[\prob f]\right) - N\left(H[\maxentP] - \langle H[\prob p]\rangle \right)
= 
N \infdiv{\prob f}{\maxentP} 
- (\dimstatespace-\widehat D)
=
\sfrac12\widehat{\textsc{aic}} - NH[\prob f] - \dimstatespace
~.
}
In the spirit of \eqref{eq:Motivation:ExponentialDominanance}, minimization of \textsc{aic} score (Akaike information criterion~\cite{akaike1974new}) can be understood as selecting the set of constraints whose \maxent-to-expected-entropy ratio of combinatorial possibilities is larger than the expected-to-empirical-entropy ratio.

\section{Training and test metrics}

For the purposes of this section, we suppose that some form of ``truth'' asymptotically exists.
Let $\prob q\in\mathcal P$ describe the asymptotic model as  some \maxent distribution of architecture $\coefM$. 
To assess the quality of modeling at a given sample size $N$ we first define the expected training error of an architecture $\widehat{\coefM}$ by
\equ{
\label{eq:TrainingError}
\varepsilon_\text{train} = \sum_{\prob f\in\mathcal P} \multidistro(N\prob f; \prob q)\, N \infdiv{\prob q}{\maxentP}
~,
}
where $\maxentP\in[\prob f]_{\widehat{\coefM}}$ is understood to be the \maxent distribution in the equivalence class induced by $\widehat{\coefM}$ on samples $\prob f$ drawn from $\prob q$. 
This error is governed by the expected Kullback-Leibler (\textsc{kl}) divergence~\cite{kullback1997information} of model from sample distribution for samples drawn from the ``true'' distribution $\prob q$.
Intuitively, the training error quantifies the average proximity to $\prob q$ achieved by models $\maxentP$ that were  trained on a sample of size $N$. It depends on $N$, the training architecture and the unknown asymptotic model.
To evaluate \eqref{eq:TrainingError} we consider two distinct cases. 

If  model architecture $\widehat{\coefM}$ implies all asymptotic constraints $\coefM$, then there must exist parameter vector $\hat{\boldsymbol\theta}\in\mathbb R^{\hat D}$ such that 
\equ{
\prob q_\ga = \exp \left(\sum_{a=1}^{\widehat D}\hat\theta_a \widehat{\coef}_{a\ga}\right)
~.
}
As data varies,  linearized fluctuations of model distribution $\maxentP$ around $\prob q$,
\equ{
\label{eq:TrainError:ModelDistro}
\maxentP_\ga = \prob q_\ga \left(1 + \frac{1}{\sqrt N}\sum_{a=1}^{\widehat D} y_a \widehat{\coef}_{a\ga} \right)
}
(column-wise multiplication by $\prob q_\ga\neq0$ is a rank-preserving operation)
induce a parametrization of some generic distribution $\prob f$ around the asymptotic distribution $\prob q$, 
\equ{
\label{eq:Genericfluctuations:Param}
\prob f_\ga = \maxentP_\ga
+ \frac{1}{\sqrt N}\sum_{i=1}^{\dimstatespace-\widehat D} x_i \widehat{X}_{i\ga}
=
\prob q_\ga + \frac{1}{\sqrt N}\left[\sum_{a=1}^{\widehat D} y_a \left(\widehat{\coef}_{a\ga}\prob q_\ga \right)
+ \sum_{i=1}^{\dimstatespace-\widehat D} x_i \widehat{X}_{i\ga}\right]
~,
}
where $\widehat{\mathbf X}$ orthogonally encodes  the kernel of $\widehat{\coefM}$, similarly to \eqref{eq:OrthonormalKernel}. Note in particular that normalization condition on the simplex $\mathcal P$ realized by $\bar a$-th row in the coefficient matrix with  $\widehat C_{\bar a\ga}=1$ automatically ensures 
\equ{
\label{eq:TrainError:xCondition}
\sum_{\ga=1}^\dimstatespace X_{i\ga} = \sum_{\ga=1}^\dimstatespace \widehat C_{\bar a\ga} \widehat X_{i\ga}= \sum_{a=1}^{\widehat D}(\mathbf T^{-1})_{\bar aa} \underbrace{\sum_{\ga=1}^\dimstatespace \widehat \coef_{a\ga} \widehat X_{i\ga}}_{=0} = 0
\quad\forall\, i=1,\ldots\dimstatespace-\widehat D~,
}
where $\mathbf T$ represents the Gaussian-Jordan transformation matrix from $\widehat{\mathbf C}$ to $\widehat{\coefM}$.
Enforcing thus $\prob f\in\mathcal P$ requires  
that the components of $y\in\mathbb R^{\widehat D}$ are linearly dependent,
\equ{
\label{eq:TrainError:yCondition}
\sum_{a=1}^{\widehat D} y_a \underbrace{\sum_{\ga=1}^\dimstatespace\widehat{\coef}_{a\ga} \prob q_\ga}_{\text{``true'' $\widehat{\prob m}$}} = 0
}
by vanishing projection on the moments of $\widehat{\coefM}$ induced on the ``true'' distribution.
Hence, there remain $\widehat D-1$ directions along which model distribution $\maxentP$ can vary.

At large $N$, expansion \eqref{eq:multidistro:expansion_LargeN} states that multinomial distribution is governed by the ``distance'' of asymptotic $\prob q$ to empirical distribution $\prob f$. Upon using \eqref{eq:Genericfluctuations:Param} the information-theoretic approximation  to multinomial distribution gives a leading quadratic (in the fluctuations) term,
\equ{
\label{eq:TrainError:GaussianKernel}
-\log\multidistro(N\prob f; \prob q) = 
\sfrac12 \sum_{a,b=1}^{\widehat D} y_a y_{b} \sum_{\ga=1}^\dimstatespace \widehat{\coef}_{a\ga} \prob q_\ga\widehat{\coef}_{b\ga}
+
\sfrac12 \sum_{i,j=1}^{\dimstatespace-\widehat D} x_i x_{j} \sum_{\ga=1}^\dimstatespace \widehat{X}_{i\ga} \frac{1}{\prob q_\ga}\widehat{X}_{j\ga}
+ \order{\log N}
~.
}
As argued in \eqref{eq:InvertibleJacobian}, precision matrices of the above form are invertible, as long as $\coefM$ and $\mathbf X$ are of full row-rank.
Any linear term in the fluctuations vanishes identically due to normalization condition \eqref{eq:TrainError:xCondition} or drops after imposing \eqref{eq:TrainError:yCondition},  whereas mixed terms in $x$ and $y$ vanish due to the normality of $\widehat{\mathbf X}$ space to $\widehat{\mathbf R}$.
This large-$N$ expansion of multinomial distribution shows that it asymptotically goes into a multivariate normal distribution, thus 
enabling us to nicely write the integral representation of \eqref{eq:TrainingError}, which takes  the form of an $(\dimstatespace-1)$-dimensional integral in $x$ and $y$.
In particular, the Gaussian kernel associated to \eqref{eq:TrainError:GaussianKernel} can be always diagonalized by an orthogonal transformation in the fluctuations.
At the same time, the integrand becomes 
\equ{
N \infdiv{\prob q}{\maxentP} = \sfrac12 \sum_{a,b=1}^{\widehat D} y_a y_{b} \sum_{\ga=1}^\dimstatespace \widehat{\coef}_{a\ga} \prob q_\ga\widehat{\coef}_{b\ga}
+ \order{N^{-1/2}}
~,
}
rendering straight-forward the computation of 
\equ{
\label{eq:TrainError_evaluated}
\varepsilon_\text{train} = \frac{\widehat D - 1}{2} + \order{N^{-1/2}}
~.
}
Whenever our model fully captures the asymptotic architecture, the expected error converges as $N\rightarrow\infty$. 
It is optimal precisely when $\widehat{\coefM} = \coefM$.

On the other hand, if $\widehat{\coefM}$ misses any of the constraints from asymptotic architecture $\coefM$, 
the large-$N$ localization of the integral representation of \eqref{eq:TrainingError} at the saddle point would give
\equ{
\varepsilon_\text{train} =  N \infdiv{\prob q}{\maxentP} + \order{\sqrt{N}}
~,
}
where $\maxentP$ is the model distribution induced by $\widehat{\coefM}$ on the asymptotic counts $N\prob q$. Since the \textsc{kl} divergence would be finite --\,model architecture lacking some constraints\,-- the training error diverges linearly in $N$ in this case.

Similarly, one can estimate the expected test (or generalization) error of a model architecture $\widehat{\coefM}$,
\equ{
\label{eq:TestError}
\varepsilon_\text{test} = \sum_{\prob f, \prob g\in\mathcal P} \multidistro(N\prob g; \prob q) \multidistro(N\prob f; \prob q) N \infdiv{\prob g}{\maxentP}
~,
}
where $\maxentP\in[\prob f]_{\widehat{\coefM}}$ represents the model distribution, as before. Parameterizing fluctuations around the asymptotic distribution $\prob q$ along the lines of \eqref{eq:TrainError:ModelDistro} and \eqref{eq:Genericfluctuations:Param}, we obtain
\equ{
\label{eq:TestError_evaluated}
\varepsilon_\text{test} = \frac{\dimstatespace + \widehat D - 2}{2} + \order{N^{-1/2}}~,
}
if test model $\widehat{\coefM}$ implies the asymptotic architecture $\coefM$ and an estimate that diverges as $N\rightarrow\infty$, whenever any of the asymptotic constraints is missed by $\widehat{\coefM}$.

\section{Model selection in the inverse Ising problem}

A classical paradigm used both in benchmark tests and to illustrate various properties of magnetic materials is the Ising model.
Suppose that a group of Ising spins $\sigma_i\in\{0,1\}$ in lattice-gas gauge  labeled by site index $i=1,\ldots L$, which could have been e.g.\ extracted from a much bigger Ising crystal, had energy (Hamiltonian)
\equ{
\label{eq:IsingHamiltonian}
E(\vec\sigma) = 
\sum_{i=1}^L h_i \sigma_i
+
\sfrac12\sum_{i,j=1}^L J_{ij} \sigma_i \sigma_j
+
\sfrac16\sum_{i,j,k=1}^L T_{ijk} \sigma_i \sigma_j \sigma_k
~.
}
As we marginalize over lattice regions, there is nothing that prevents us from writing effective associations of higher orders among the remaining spins, but for concreteness we consider up to cubic interactions. Furthermore, we take $L=5$ to speed up evaluation of multiple instances tested in parallel.
For our benchmark tests below, we shall randomly generate different values for bias vector $\mathbf h$, symmetric coupling matrix $\mathbf J$ and cubic tensor $\mathbf T$ (excluding self-interactions which are presently meaningless) based on the hypergraph 
\equ{
\label{eq:IsingGraph}
\mathcal G_\text{Ising}=\{\{1,2,3\}, \{1,2,4\}, \{3,5\}, \{4,5\}\}
~.
}
For simple Ising systems, any $n$-point feature association implies all lower terms. For example, a trilinear association $\{i,j,k\}$ automatically induces all pairwise associations among spins  $\{i,j\}$, $\{i,k\}$ and $\{j,k\}$ as well as the one-site biases for $i$, $j$ and $k$. Hence, the Ising system has $5$ non-vanishing biases, $7$ couplings and 2 cubic interactions.
The Boltzmann distribution 
\equ{
\prob q(\vec\sigma) = \left(\sum_{\sigma_1'=0,1}\cdots\sum_{\sigma'_L=0,1}e^{E(\vec\sigma')}\right)^{-1} e^{E(\vec\sigma)}
} 
of statistical system~\eqref{eq:IsingHamiltonian} defines an asymptotic model  with\footnote{In the parametric formulation of thermodynamics, the normalization condition on $\mathcal P$ is realized  by the partition function.} $D=15$ in a spin space of $\dimstatespace=2^5=32$ microstates.

From this Ising system in the inverse problem, a modeler obtains samples at different sizes $N$. Knowing only that this is a system of potentially interacting spins leaves $7\,580$ candidate hypergraphs with distinct architectures $\widehat{\coefM}$ reflecting different interaction structures.  For each architecture, \ipf or any Newton-based method can quickly deduce the corresponding \maxent distribution on a given training sample, as outlined in Section~\ref{sc:IterativeMethods}. In such setting,  applications  are usually concerned with two intimately related aspects: (i) the reconstruction of the original hypergraph $\mathcal G_\text{Ising}$ from a sample of size $N$ and (ii) the generalization error of a trained model on different samples  drawn from the same system.

To answer those questions by better understanding the various information-theoretic metrics discussed in Section~\ref{sc:ConstraintSelection}, 
we have independently generated 10\,000 samples from $1\,000$ different realizations of the Ising system \eqref{eq:IsingHamiltonian}  with interaction structure \eqref{eq:IsingGraph} at various sample sizes.
By a realization, we simply mean a set of fields and couplings parametrizing the Ising Hamiltonian, which were independently drawn from a normal distribution of zero mean and $\order{1}$ variance.
On each training dataset of size $N$ the \maxent distribution associated to all $7\,580$ architectures is deduced. Subsequently, the architecture $\widehat \coefM$ is recorded each time which exhibits
(a) lowest \textsc{bic},
(b) lowest \textsc{aic}, 
(c) lowest $\widehat D$ \textit{and} empirical $p$-value above threshold (hyper-\maxent),
(d) lowest $\widehat D$ \textit{and} typical for all more complex models implying $\widehat \coefM$ (hyper-\maxent + \textsc{lrt}).

In order to determine a threshold for (c) and (d) below which a  $p$-value is considered to be significant (see Section~\ref{sc:HypothesisTesting}) 
we exploit two facts.  
According to \eqref{eq:TrainError_evaluated} and \eqref{eq:TestError_evaluated}, expected errors tend to decrease for simpler models, hence the cutoff should increase for higher-dimensional kernels \eqref{eq:OrthonormalKernel} for which we anticipate on average smaller atypical fluctuations.  
On the other hand, larger datasets tend to eliminate sampling noise sharpening the distribution of $p$-values. The empirical $p$-value associated to any architecture that misses some asymptotic constraint goes faster to zero, while any slightly atypical dataset sampled from $\prob q$ could lead to smaller $p$-values under the asymptotic model. 
The acceptance threshold should thus scale with inverse sampling size to ensure a proper asymptotic limit. 
Combining these observations we conclude that the fraction
\equ{
\label{eq:Threshold1}
\alpha_\text{empirical} = \frac{\dimstatespace-\widehat D}{N}
}
has a  scaling which seems reasonable for a cut-off in the \maxent hypothesis testing.
In a similar spirit, 
we use  in the \textsc{lrt} part of (d) as  significance level  
\equ{
\label{eq:Threshold2}
\alpha_\textsc{lrt} = \frac{2\dimstatespace-\widehat D-\widehat D'}{N}
}
to decide about the $p$-value when comparing two nested model architectures $\widehat\coefM$ and $\widehat\coefM'$.

\begin{figure}
\centering
\includegraphics[scale=0.51]{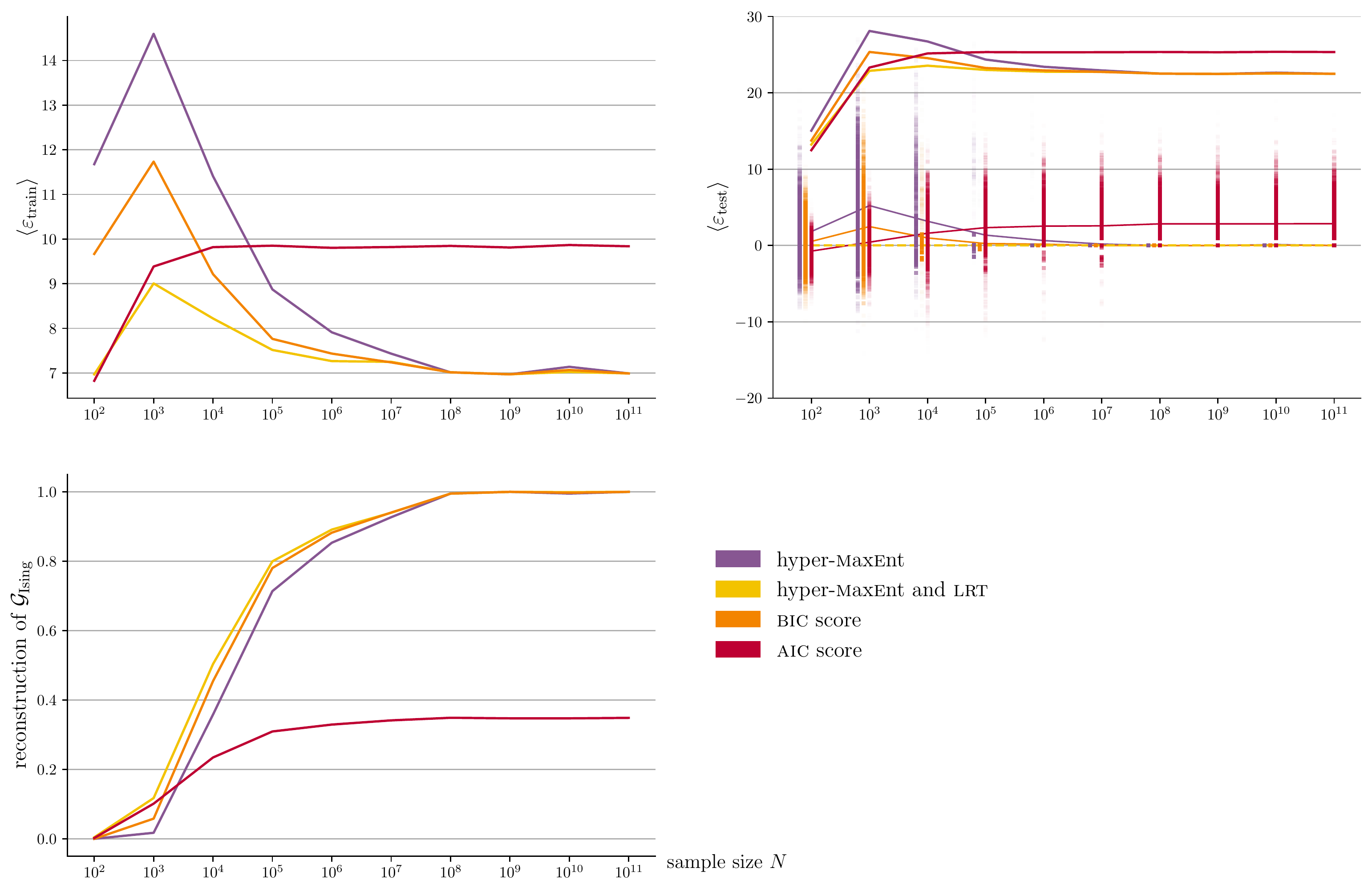}
\caption{\small{Left: Average ``distance'' from  the asymptotic distribution associated to $\mathcal G_\text{Ising}$ of model distribution  that is selected by the various scores as a function of sample size $N$. Below: Average reconstruction accuracy determined by the number of correctly identified hypergraphs $\mathcal G_\text{Ising}$ in the generated samples. Right: Average generalization error (and its dispersion) of model distribution on unseen data sampled from the asymptotic system as a function of $N$.}}\label{fg:BenchmarkIsing}
\end{figure}
\begin{figure}
\centering
\includegraphics[scale=0.35]{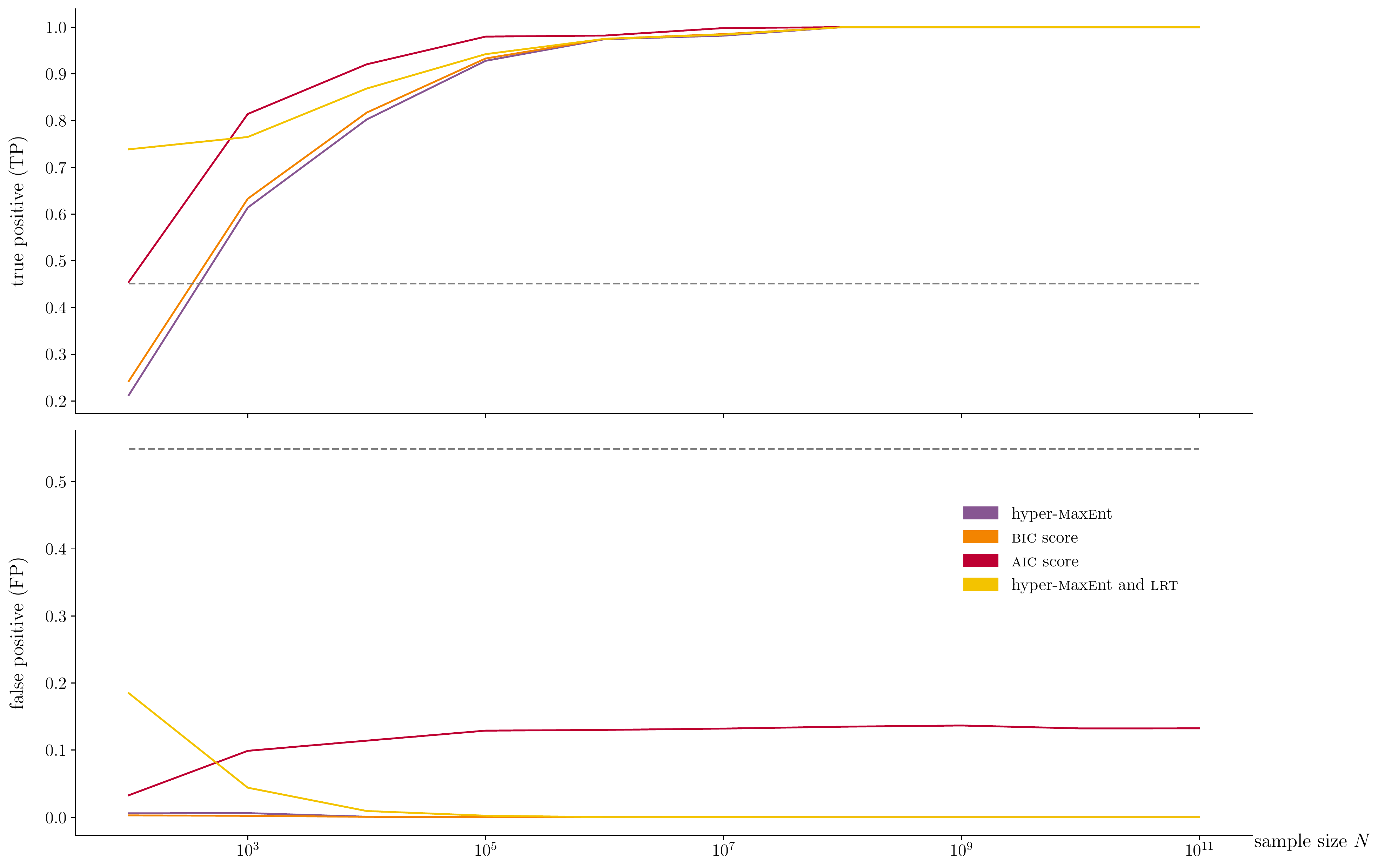}
\caption{\small{$N$-dependent average rates of true/false positive associations against asymptotic graph $\mathcal G_\text{Ising}$ characterizing the various selection procedures.}}\label{fg:ROC}
\end{figure}
In Figure~\ref{fg:BenchmarkIsing}, we plot the training error \eqref{eq:TrainingError} averaged over all sampled datasets of various  sizes $N=10^2,10^3,\ldots, 10^{11}$. To get a better idea on the training quality the average graph reconstruction that was achieved at each sample size is comparatively plotted. This shows how often model  architecture  $\widehat \coefM$ selected by the various metrics precisely captures the underlying Ising model associated to graph \eqref{eq:IsingGraph}, but nothing more.
On the right plot, the test error \eqref{eq:TestError} on $1\,000$ samples is given again averaged over the $10^7$ training datasets at each sample size. To obtain a feeling about the dispersion of the generalization error we alongside plot the difference of the test error of the other scores from the consistently best-scoring \textsc{lrt} model within each sample. 

We observe that hyper-\maxent and \textsc{lrt} with the acceptance threshold introduced in \eqref{eq:Threshold1} and \eqref{eq:Threshold2} as well as the \textsc{bic} score are all scoring metrics that are consistent with large-$N$ expansion(s). As $N$ increases, they clearly exhibit the correct asymptotic behavior eventually achieving $100\%$ reconstruction accuracy. At the same time, the generalization error saturates the bound in \eqref{eq:TestError} for $\widehat D=D$. 
It is important to stress that modifying the significance level, does not change the asymptotic behavior, as long as the scaling goes (at least) as $1/N$. This demonstrates the universality of large-$N$ expansion which renders any compatible selection algorithms indistinguishable as $N$ tends to infinity. 
On the other hand, \textsc{aic} consistently appears sub-optimal failing to recover the underlying model. Instead, the reasoning of maximizing the expected entropy \eqref{eq:ExpectedEntropy} seems to capture models that are generically more complex than necessary and are thus plagued by false positive (\textsc{fp}) associations.

To get a handle on true/false positive (\textsc{tp}/\textsc{fp}) rates, we plot\footnote{A Python script to randomly generate Ising systems of a given architecture, sample from them and benchmark the presented model-selection procedures can be found under \href{https://github.com/imbbLab/ModelSelection}{github.com/imbbLab/ModelSelection}.} the percentages of correctly identified interactions, which are achieved by the various selection procedures, averaged over all sampled datasets at each size $N$. As a base reference, we use the probability that a randomly selected constraint out of the 31 possible  is a true (false) positive, i.e.\ falls (does not fall) within the set of the 14 asymptotic constraints. This \textsc{roc}-curve inspired plot verifies our previous finding on the large-$N$ compatibility of all three selection procedures except for the \textsc{aic} score.

However, it is noteworthy that \textsc{aic} manages to quickly reach a higher \textsc{tp} rate at the expense of keeping on average a finite \textsc{fp} rate, which in the present setting remains lower than the base rate (dotted gray line).
By modifying acceptance  threshold \eqref{eq:Threshold2}, one can adjust the \textsc{fp} tolerance of \textsc{lrt} selection in the small-$N$ regime.
Most interestingly, we recognize that \textsc{bic} and the hyper-\maxent consistently have (almost) vanishing \textsc{fp} rates, even at sample sizes of order $100$. In particular, this gives a strong hint in favor of the $\order{1/N}$ scaling of acceptance threshold \eqref{eq:Threshold1}. To further support the suggested scaling for the threshold, we mention the observation that the architecture $\widehat\coefM=\coefM$ corresponding to the asymptotic graph \eqref{eq:IsingGraph} always has an empirical $p$-value \eqref{eq:empirical_p_value} above $\ga_\text{empirical}$ at any sample size $N$.

\bibliographystyle{paper}
{\small
\bibliography{paper}

\providecommand{\href}[2]{#2}\begingroup\raggedright\begin{thebibliography}{10}

\bibitem{6773024}
C.~E. Shannon ``A mathematical theory of communication'' {\em The Bell System
  Technical Journal} {\bf 27} (1948) no.~3, 379--423.

\bibitem{jaynes2003probability}
E.~T. Jaynes {\em Probability theory: The logic of science}.
\newblock Cambridge university press 2003.

\bibitem{shore1980axiomatic}
J.~Shore and R.~Johnson ``Axiomatic derivation of the principle of maximum
  entropy and the principle of minimum cross-entropy'' {\em IEEE Transactions
  on Information Theory} {\bf 26} (1980) no.~1, 26--37.

\bibitem{paris1990note}
J.~B. Paris and A.~Vencovsk{\'a} ``A note on the inevitability of maximum
  entropy'' {\em International Journal of Approximate Reasoning} {\bf 4} (1990)
  no.~3, 183--223.

\bibitem{csiszar1996maxent}
I.~Csisz{\'a}r ``Maxent, mathematics, and information theory'' in {\em Maximum
  entropy and Bayesian methods} pp.~35--50.
\newblock Springer 1996.

\bibitem{keynes1921chapter}
J.~M. Keynes ``Chapter iv: The principle of indifference'' {\em A treatise on
  probability} {\bf 4} (1921) 41--64.

\bibitem{loukas2022categorical}
O.~Loukas and H.~R. Chung ``Categorical distributions of maximum entropy under
  marginal constraints'' {\em arXiv preprint arXiv:2204.03406} (2022).

\bibitem{jaynes1968prior}
E.~T. Jaynes ``Prior probabilities'' {\em IEEE Transactions on systems science
  and cybernetics} {\bf 4} (1968) no.~3, 227--241.

\bibitem{csiszar1975divergence}
I.~Csisz{\'a}r ``I-divergence geometry of probability distributions and
  minimization problems'' {\em The annals of probability} (1975) 146--158.

\bibitem{amari2000methods}
S.-i. Amari and H.~Nagaoka {\em Methods of information geometry} vol.~191.
\newblock American Mathematical Soc. 2000.

\bibitem{geyer2013asymptotics}
C.~Geyer and G.~Meeden ``Asymptotics for constrained dirichlet distributions''
  {\em Bayesian Analysis} {\bf 8} (2013) no.~1, 89--110.

\bibitem{Rosenkrantz1989}
E.~T. Jaynes {\em Concentration of Distributions at Entropy Maxima (1979)}
  pp.~315--336.
\newblock Springer Netherlands Dordrecht 1989.

\bibitem{ireland1968contingency}
C.~T. Ireland and S.~Kullback ``Contingency tables with given marginals'' {\em
  Biometrika} {\bf 55} (1968) no.~1, 179--188.

\bibitem{bishop2007discrete}
Y.~M. Bishop, S.~E. Fienberg, and P.~W. Holland {\em Discrete multivariate
  analysis: theory and practice}.
\newblock Springer Science \& Business Media 2007.

\bibitem{silvey1975statistical}
S.~Silvey {\em Statistical Inference}.
\newblock Chapman \& Hall/CRC Monographs on Statistics \& Applied Probability.
  Taylor \& Francis 1975.

\bibitem{doi:10.1080/00031305.1982.10482817}
A.~Buse ``The likelihood ratio, wald, and lagrange multiplier tests: An
  expository note'' {\em The American Statistician} {\bf 36} (1982) no.~3a,
  153--157
  \href{http://www.arXiv.org/abs/https://doi.org/10.1080/00031305.1982.10482817}{[{\tt
  arXiv:https://doi.org/10.1080/00031305.1982.10482817}]}.

\bibitem{10.1214/aos/1176344136}
G.~Schwarz ``{Estimating the Dimension of a Model}'' {\em The Annals of
  Statistics} {\bf 6} (1978) no.~2, 461 -- 464.

\bibitem{akaike1974new}
H.~Akaike ``A new look at the statistical model identification'' {\em IEEE
  transactions on automatic control} {\bf 19} (1974) no.~6, 716--723.

\bibitem{kullback1997information}
S.~Kullback {\em Information theory and statistics}.
\newblock Courier Corporation 1997.

\end{thebibliography}\endgroup
}

\end{document}